\documentclass[conference,10pt,letterpaper]{IEEEtran}
\IEEEoverridecommandlockouts
\usepackage{amsmath,amssymb,bm}
\usepackage{mathtools}
\usepackage{graphicx,cite,booktabs,etoolbox}
\usepackage[caption=false,font=footnotesize]{subfig}
\makeatletter
\patchcmd{\@makecaption}{\hfil\box\@tempboxa\hfil}{\box\@tempboxa\hfil}{}{}
\patchcmd{\@makecaption}{\normalfont\footnotesize\noindent\unhbox}
    {\normalfont\footnotesize\raggedright\noindent\unhbox}{}{}
\makeatother
\usepackage[ruled,linesnumbered]{algorithm2e}
\SetAlgoSkip{medskip}
\usepackage[hidelinks]{hyperref}
\allowdisplaybreaks[1]
\graphicspath{{figures/}}
\newcommand{\Tr}{\operatorname{Tr}}

\newcommand{\norm}[1]{\left\|#1\right\|}
\newtheorem{lemma}{\bf{Lemma}}
\begin{document}
\bstctlcite{BSTcontrolFirstAuthor}
\title{\LARGE Robust Beamforming and Antenna Position Optimization for Movable Antenna-Assisted Covert Communications}
\author{\IEEEauthorblockN{Xiangyu Dong\IEEEauthorrefmark{1}, Ran Yang\IEEEauthorrefmark{1}, Weidong Mei\IEEEauthorrefmark{1}, Zheng Dong\IEEEauthorrefmark{2}, Yue Xiu\IEEEauthorrefmark{1}, Zhongpei Zhang\IEEEauthorrefmark{3}, and Tony Quek\IEEEauthorrefmark{4}}
\IEEEauthorblockA{\IEEEauthorrefmark{1}\textit{National Key Laboratory of Wireless Communications,}\\
\textit{University of Electronic Science and Technology of China, Chengdu, China}}
\IEEEauthorblockA{\IEEEauthorrefmark{2}\textit{School of Information Science and Engineering, Shandong University, Qingdao, China}}
\IEEEauthorblockA{\IEEEauthorrefmark{3}\textit{Shenzhen Institute for Advanced Study, University of Electronic Science and Technology of China, Shenzhen, China}}
\IEEEauthorblockA{\IEEEauthorrefmark{4}\textit{Information Systems Technology and Design, Singapore University of Technology and Design, Singapore}}
\IEEEauthorblockA{Email: zhangzp@uestc.edu.cn}
\IEEEauthorblockA{\textit{Corresponding Author: Zhongpei Zhang}
\thanks{This work was supported in part by the National Key Research and Development Program of China under Grant 2024YFB2907900, and in part by Shenzhen Natural Science Foundation under Grant JCYJ20250604180459039.}}}
\maketitle
\begin{abstract}
In this paper, we study a movable antenna (MA)-enhanced covert communication system, where a legitimate transmitter (Alice) serves multiple legitimate users in the presence of multiple non-colluding Willies that attempt to detect the covert transmission from Alice.
To ensure data covertness with imperfect knowledge of the Willies' channel state information (CSI), we develop a robust covert transmission framework with a norm-bounded CSI error model.
In particular, we aim to jointly optimize the transmit beamforming vectors, artificial noise (AN) covariance matrix, and MA positions to maximize the weighted minimum signal-to-interference-plus-noise ratio (SINR) among all legitimate users subject to worst-case covertness constraints.
To tackle this challenging problem, we develop a block coordinate descent (BCD)-based optimization framework, incorporating the S-procedure, semidefinite relaxation (SDR), bisection search, and successive convex approximation (SCA).
Numerical results demonstrate the effectiveness and robustness of the proposed
algorithm in improving the performance of MA-enhanced covert communication systems.
\end{abstract}
\begin{IEEEkeywords}
Movable antenna, covert communications, robust beamforming, artificial noise.
\end{IEEEkeywords}
\section{Introduction}\label{sec:intro}
The sixth-generation (6G) mobile networks are envisioned to provide ubiquitous connectivity for emerging applications \cite{Vision6G}.
Conventional wireless networks primarily serve terrestrial devices, whereas 6G networks are expected to integrate terrestrial and aerial networks.
In particular, the low-altitude economy (LAE) is driving the deployment of a growing number of uncrewed aerial vehicles (UAVs).
These UAVs can be integrated into cellular networks as aerial users~\cite{aerialuser} or aerial sensing platforms~\cite{MAISAC2} for applications such as surveillance, data collection, and aerial filming~\cite{11196017}.
However, the broadcast nature of wireless channels exposes low-altitude transmissions to security risks such as eavesdropping, information leakage, and jamming \cite{AlHabobAerialPLS,XiuMAAntiJammingMEC}. 
These risks underscore the fundamental need for reliable and secure wireless transmission in low-altitude networks~\cite{citeformeipls}.

In this context, covert communication has drawn wide
attention for its capability to conceal transmission behavior \cite{WangPLSCovert}.
The authors in \cite{Bash2013} pioneered to establish the fundamental square root law, which states that $\mathcal{O}(\sqrt{N})$ bits can be delivered reliably and covertly over $N$ uses of an additive white Gaussian noise (AWGN) channel.
In \cite{SobersJammer2017}, an uninformed jammer transmitting artificial noise (AN) was proposed to increase the number of reliably and covertly transmitted bits from $\mathcal{O}(\sqrt{N})$ to $\mathcal{O}(N)$.
In particular, multi-antenna techniques have been widely adopted in covert communications by offering spatial degrees of freedom (DoFs) in both detection and transmission.
From the detection perspective, the authors in \cite{HuBeamSweeping} derived the optimal number of beam-sweeping sectors that minimizes the detection error probability (DEP) of a UAV transmission.
However, most existing designs rely on fixed-position antenna (FPA) arrays, leaving the spatial DoFs associated with antenna positions unexploited and limiting covertness improvement.

Recently, movable antennas (MAs) have emerged as an innovative solution to overcome this limitation by dynamically adjusting the antenna positions within a wavelength-scale region~\cite{MA4,YangFlexibleWMMSE2025}.
Unlike conventional FPA arrays, MA arrays can avoid deep fading and reshape the composite channel through changes in the propagation phases of the multipath components with antenna repositioning~\cite{MA1}.
The spatial flexibility of MAs has also been exploited in covert communication.
The authors in \cite{WangWavelengthCovert} leveraged antenna repositioning to achieve wavelength-scale beamforming resolution, preserving covertness even with a Willie near the intended receiver.
In \cite{LiuMACovert}, the transmit beamforming and MA positions were jointly optimized for a covert user in the presence of multiple non-colluding Willies.
Although these works have demonstrated the effectiveness of MAs, they are typically based on the perfect knowledge of the Willie channel state information (CSI).
In practice, perfect CSI of non-cooperative Willies is challenging for Alice to acquire~\cite{citeformeirobust,CovertSurvey}.
Therefore, ignoring the inherent CSI error may violate the covertness requirement, leading to degraded security performance.
To address this issue,
the authors in \cite{YangMAISACC2026} investigated robust beamforming and antenna position optimization for covert ISAC systems.
However, the proposed approach in \cite{YangMAISACC2026} was limited to scenarios with a single Willie.
To enhance data security and further unlock the synergy gain between MAs and covert transmission, a robust MA-based transmission framework is highly imperative.

In this paper, we study robust MA-assisted covert communications with multiple Willies under imperfect CSI. The main contributions are summarized as follows.
1) To guarantee transmission covertness, we develop a worst-case optimization framework to maximize the weighted max-min signal-to-interference-plus-noise ratio (SINR) by jointly optimizing transmit beamforming, AN covariance matrix, and antenna placement, subject to a transmit power budget and a covertness constraint at each Willie. 
2) To address the challenging problem, we first apply the S-procedure to reformulate the intractable semi-infinite quadratic condition into an equivalent linear matrix inequality (LMI). 
Then, we develop a block coordinate descent (BCD) algorithm combining semidefinite relaxation (SDR), bisection search, and successive convex approximation (SCA) methods.
3) Simulation results show that the proposed MA schemes achieve superior performance in weighted max-min SINR compared with the conventional FPA schemes, with further improvements observed for larger movable regions and richer multipath propagation.
\section{System Model and Problem Formulation}
\begin{figure}[t]
\centering
\includegraphics[width=0.8\columnwidth]{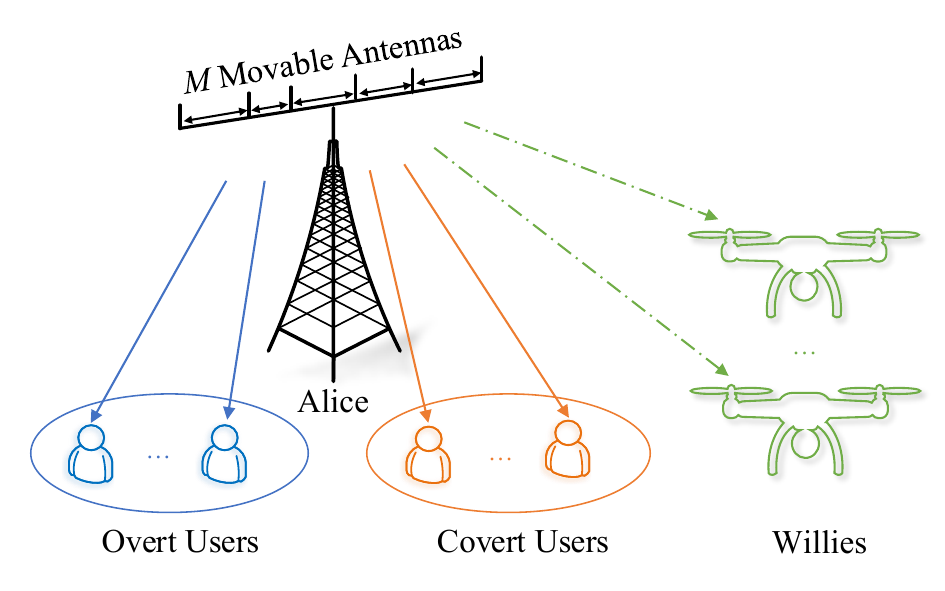}
\caption{MA-assisted covert communication system.}
\label{fig:system_model}
\end{figure}

We consider an MA-assisted covert communication system, as shown in Fig.~\ref{fig:system_model}, where Alice serves $K_o$ overt users and $K_c$ covert users in the presence of $Q$ Willies, with $K=K_o+K_c$.
Alice is equipped with $M$ MAs for signal transmission, while each legitimate user and Willie is equipped with a single antenna.
For brevity, $\mathcal K_o=\{1,\ldots,K_o\}$ and $\mathcal K_c=\{K_o+1,\ldots,K_o+K_c\}$ denote the index sets of the overt and covert users, respectively, with $\mathcal K=\mathcal K_o\cup\mathcal K_c$.
The index set of the Willies is denoted by $\mathcal Q=\{1,\ldots,Q\}$.
The Willies are assumed to be non-colluding and independently attempt to detect whether Alice is transmitting confidential information to the covert users or not.
We exploit the overt signals and AN as a cover to achieve covert communications.
It is assumed that the feasible movement range for the transmitting MAs is a one-dimensional (1D) interval of length $D$.
The transmitting antenna positioning vector (APV) is denoted by $\bm t=[t_1,\ldots,t_M]^T$, with $0\le t_1\le t_2\le\cdots\le t_M\le D$.

\subsection{Signal Model}
Denote by $\mathcal H_0$ and $\mathcal H_1$ the hypotheses corresponding to the absence and presence of covert transmission, respectively. The transmitted signal at Alice in time slot $n$ is given by
\begin{equation}\label{eq:tx_signal}
\left\{
\begin{aligned}
\mathcal{H}_1:\bm{s}[n] &= \bm W_o\bm c_o[n]+\bm W_c\bm c_c[n]+\bm r[n],\\
\mathcal{H}_0:\bm{s}[n] &= \bm W_o\bm c_o[n]+\bm r[n],
\end{aligned}
\right.
\end{equation} 
where $\bm W_o=[\bm w_1,\ldots,\bm w_{K_o}]\in\mathbb C^{M\times K_o}$ is the overt beamforming matrix and $\bm W_c=[\bm w_{K_o+1},\ldots,\bm w_{K_o+K_c}]\in\mathbb C^{M\times K_c}$ is the covert beamforming matrix.
The overt data vector is $\bm c_o[n]=[c_1[n],\ldots,c_{K_o}[n]]^T\in\mathbb C^{K_o\times1}$, while the covert data vector is $\bm c_c[n]=[c_{K_o+1}[n],\ldots,c_{K_o+K_c}[n]]^T\in\mathbb C^{K_c\times1}$.
We consider transmitting an AN vector $\bm r[n]$ to interfere with the Willies.
It is assumed that the data symbols $c_k[n]$, $k\in\mathcal K$, and the AN vector $\bm r[n]$ are mutually independent, with $c_k[n]\sim\mathcal{CN}(0,1)$ and $\bm r[n]\sim\mathcal{CN}(\bm0,\bm R_{\rm A})$.

\subsection{Channel Model}
Given that the propagation distance is significantly larger than the size of the moving regions, the far-field response is adopted for channel modeling~\cite{zhu2025tutorial,P_FAA}.
To be specific, angle-of-departure (AoD), the angle-of-arrival (AoA) and amplitude of the complex coefficient for each link remain constant independent of MA positions.
It is assumed that the geometric model is adopted for the communication channels, where the number of transmission paths is the same for all Alice--user links, denoted by $L$~\cite{MA1}.
Consequently, the field response vector (FRV) at $t_m$ is given by
$\bm g_k(t_m)=\left[e^{\jmath\frac{2\pi}{\lambda}t_m\cos\theta_k^1},\ldots,e^{\jmath\frac{2\pi}{\lambda}t_m\cos\theta_k^{L}}\right]^T\in\mathbb C^{L\times1}$,
where $\lambda$ denotes the carrier wavelength, and $\theta_k^\ell$ is the AoD of the $\ell$-th path.
The corresponding field response matrix (FRM) from Alice to user $k$ is given by
$\bm G_k(\bm t)=[\bm g_k(t_1),\ldots,\bm g_k(t_M)]\in\mathbb C^{L\times M}$.
Then, the channel vector at the $k$-th user is given by
\begin{equation}
\bm h_k^H(\bm t)=\bm f_k^H\bm G_k(\bm t)\in\mathbb C^{1\times M},\quad\forall k\in\mathcal K,
\label{eq:user_channel}
\end{equation}
where $\bm f_k=[\beta_k^1,\ldots,\beta_k^L]^H$ with $\beta_k^\ell$ denoting the complex path gain of the $\ell$-th path. 
Assuming that each Willie is deployed as a UAV, the nominal line-of-sight (LoS) channel from Alice to Willie $q$ is given by
\begin{equation}
\bar{\bm h}_q^H(\bm t)
=\beta_q\bm g_q(\bm t)\in\mathbb C^{1\times M},
\forall q \in \mathcal{Q},
\label{eq:warden_channel}
\end{equation}
where $\bm g_q(\bm t)=\left[e^{\jmath\frac{2\pi}{\lambda}t_1\cos\theta_q},\ldots,e^{\jmath\frac{2\pi}{\lambda}t_M\cos\theta_q}\right]\in\mathbb C^{1\times M}$, $\beta_q$ and $\theta_q$ denote the estimated LoS path gain and AoD, respectively.

\subsection{Communication Performance}
Under $\mathcal{H}_1$, the received signal at the legitimate user $k$ is given by
\begin{small}
\begin{equation}\label{eq:rx_user}
\begin{aligned}
y_k[n]\!=\!\bm h_k^H\!(\bm t)\bm w_k c_k[n]\!+\!\sum_{i\in\mathcal K\setminus\!\{\!k\!\}}\!\bm h_k^H\!(\bm t)\bm w_i c_i[n]\!+\!\bm h_k^H\!(\bm t)\bm r[n]\!+\!z_k[n],\!
\end{aligned}
\end{equation}
\end{small}%
where $z_k[n]\sim\mathcal{CN}(0,\sigma_k^2)$ denotes the additive white Gaussian noise (AWGN) at the $k$-th legitimate user.
The corresponding SINR at user $k$ under hypothesis $\mathcal H_1$ is given by
\begin{equation}
\gamma^{1}_{k}\!=\!
\frac{\bm h_{k}^{H}(\bm t)\bm W_k \bm h_{k}(\bm t)}
{\sum_{i\in\mathcal K\setminus\!\{\!k\!\}}\!\bm h_k^H(\bm t)\bm W_i \bm h_k(\bm t)\!+\!
\bm h_k^H(\bm t)\bm R_{\rm A}\bm h_k(\bm t)\!+\!\sigma_k^2},
\label{eq:sinr_user}
\end{equation}
where $\bm W_k\triangleq\bm w_k\bm w_k^H,\forall k\in \mathcal{K}$.

\subsection{Detection Performance}
We consider a non-colluding Willie scenario, where each Willie independently detects the communication behaviour between Alice and the covert users based on the signals received over $N$ time slots. Let $\bm y_q=[y_q[1],\ldots,y_q[N]]^T\in\mathbb C^{N\times1}$ denote the observation vector at Willie $q$. The received signal under the two hypotheses is expressed as
\begin{equation}\label{eq:rx_warden}
\left\{
\begin{aligned}
\mathcal H_1:y_q[n]&=\bm h_q^H(\bm t)\big(\bm W_o\bm c_o[n]+\bm W_c\bm c_c[n]+\bm r[n]\big)+z_q[n],\\
\mathcal H_0:y_q[n]&=\bm h_q^H(\bm t)\big(\bm W_o\bm c_o[n]+\bm r[n]\big)+z_q[n],
\end{aligned}
\right.
\end{equation}
where $z_q[n]\sim\mathcal{CN}(0,\sigma_q^2)$ denotes the AWGN at Willie $q$.
Assuming that the received samples are independent and identically distributed, the probability density functions (PDFs) of $\bm y_q$ under the two hypotheses are respectively given by
\begin{equation}\label{eq:warden_pdf1}
\begin{aligned}
\mathbb P(\bm y_q|\mathcal H_1)
\!=\!\frac{\exp\!\left(\!-\frac{\|\bm y_q\|^2}{\Lambda_{1,q}(\bm t)}\!\right)}{\pi^N\Lambda_{1,q}^N(\bm t)}, \mathbb P(\bm y_q|\mathcal H_0)
\!=\!\frac{\exp\!\left(\!-\frac{\|\bm y_q\|^2}{\Lambda_{0,q}(\bm t)}\!\right)}{\pi^N\Lambda_{0,q}^N(\bm t)},
\end{aligned}
\end{equation}
where $\Lambda_{1,q}(\bm t)=\bm h_q^H(\bm t)\left(\sum_{k\in\mathcal K}\bm W_k+\bm R_{\rm A}\right)\bm h_q(\bm t)+\sigma_q^2$ and $\Lambda_{0,q}(\bm t)=\bm h_q^H(\bm t)\left(\sum_{k\in\mathcal K_o}\bm W_k+\bm R_{\rm A}\right)\bm h_q(\bm t)+\sigma_q^2$.

Following the work in \cite{Ran1}, we assume that each Willie uses classical hypothesis testing with equal prior probabilities for the two hypotheses.
Let $\mathcal D_{1,q}$ and $\mathcal D_{0,q}$ represent the decisions in support of $\mathcal H_1$ and $\mathcal H_0$, respectively. The false alarm probability is given by $\mathbb P_{{\rm FA},q}=\mathbb P(\mathcal D_{1,q}|\mathcal H_0)$, while the missed detection probability is given by $\mathbb P_{{\rm MD},q}=\mathbb P(\mathcal D_{0,q}|\mathcal H_1)$.
Under optimal detection, Willie $q$ minimizes the detection error probability (DEP) $\xi_q=\mathbb P_{{\rm FA},q}+\mathbb P_{{\rm MD},q}$, with the minimum denoted by $\xi_q^*$. The covertness constraint at each Willie is expressed as
$ \xi_q^*\ge1-\epsilon,\forall q\in\mathcal Q, $
where $\epsilon\in[0,1]$ is the covertness constant.

The expression of $\xi_q^*$ involves incomplete gamma functions, making the covertness constraint difficult to handle directly.
To address this issue, we use a lower bound on $\xi_q^*$ based on the Kullback--Leibler (KL) divergence~\cite{ZhangRobustCovertISAC}.
By applying Pinsker's inequality, we have $\xi_q^*\ge1-\sqrt{\mathcal D\big(\mathbb P(\bm y_q|\mathcal H_1)\|\mathbb P(\bm y_q|\mathcal H_0)\big)/2}$. 
Therefore, the original covertness constraint is satisfied if $\mathcal D\big(\mathbb P(\bm y_q|\mathcal H_1)\|\mathbb P(\bm y_q|\mathcal H_0)\big)\le2\epsilon^2$. By substituting~\eqref{eq:warden_pdf1} into the KL divergence, we have
\begin{equation}
\begin{aligned}
\mathcal D \! \left(\mathbb P(\bm y_q|\mathcal H_1)\|\mathbb P(\bm y_q|\mathcal H_0)\right)\!=\!N\! f\!\left(\!\frac{\bm h_q^H\!(\bm t)\!\!\sum_{k\in\mathcal K_c}\!\!\bm W_{\!k}\bm h_q(\bm t)}{\Lambda_{0,q}(\bm t)}\!\right)\!,\!
\end{aligned}
\label{eq:KL_div}
\end{equation}
where $f(x)=x-\ln(1+x)$ is monotonically increasing for $x\ge0$. Hence, the sufficient condition based on the KL divergence can be equivalently rewritten as
\begin{equation}
\bm h_q^H(\bm t)\bm\Xi\bm h_q(\bm t)\le\eta\sigma_q^2,\quad\forall q\in\mathcal Q,
\label{eq:covert_compact}
\end{equation}
where $\bm\Xi=\sum_{k\in\mathcal K_c}\!\bm W_k-\eta\left(\sum_{k\in\mathcal K_o}\!\bm W_k\!+\!\bm R_{\rm A}\right)\in \mathbb{C}^{M\times M}$ and $\eta\ge0$ is the unique solution of $f(\eta)=2\epsilon^2/N$.

\subsection{Imperfect CSI Model}
In practice, the acquisition of accurate Willie CSI is challenging due to the limited cooperation between Alice and the Willies. To account for the resulting CSI uncertainty, the channel of Willie $q$ is modeled as
\begin{equation}
\bm h_q(\bm t)=\bar{\bm h}_q(\bm t)+\bm\delta_q,\quad\forall q\in\mathcal Q,
\label{eq:warden_CSI_decomposition}
\end{equation}
where $\bar{\bm h}_q(\bm t)$ is the nominal channel in~\eqref{eq:warden_channel} available at Alice, and $\bm\delta_q$ denotes the corresponding estimation error.
In this paper, we adopt a norm-bounded CSI error model, under which the estimation error satisfies
$\norm{\bm\delta_q}_2\le e_q,\forall q\in\mathcal Q,
\label{eq:bounded_CSI_error}$
where $e_q>0$ is a prescribed upper bound on the estimation error norm and is assumed known at Alice.

\subsection{Problem Formulation}
We aim to maximize the weighted minimum SINR among legitimate users by jointly optimizing the transmit beamforming vectors, the AN covariance matrix, and MA positions. Using the sufficient covertness condition~\eqref{eq:covert_compact}, we formulate the robust optimization problem as
\begin{subequations}\label{p1}
\begin{align}
\mathmakebox[2em][r]{\max_{\mathclap{\gamma,\bm W_k,\bm R_{\rm A},\bm t}}}\quad&\gamma\\
\mathmakebox[2em][r]{\text{s.t.}}\quad&\gamma^{1}_{k}\ge\nu_k\gamma,\forall k\in\mathcal K,\label{Const_1}\\
&\bm h_q^H\!(\bm t)\bm\Xi\bm h_q(\bm t)\!\le\!\eta\sigma_q^2,\norm{\bm\delta_q}_2\le e_q,\forall q\in\mathcal Q,\label{Const_2}\\
&\Tr(\bm R)\le P,\label{Const_3}\\
&t_1\ge0,\ t_M\le D,\label{Const_4}\\
&t_m-t_{m-1}\ge d,\ 2\le m\le M,\label{Const_5}
\end{align}
\end{subequations}
where $\gamma$ denotes the weighted minimum SINR threshold, $\bm R=\sum_{k\in\mathcal K}\bm W_k+\bm R_{\rm A}$, $P$ denotes the transmit power budget, $D$ and $d>0$ denote the transmit region length and minimum antenna spacing, respectively, and $\nu_k>0$ is the priority weight assigned to each user. The weights are set to $\nu_k=1$ for overt users and $\nu_k=\nu$ for covert users. Problem~\eqref{p1} is challenging to solve directly due to the infinitely many channel constraints in~\eqref{Const_2}.
\section{Proposed BCD Algorithm}
In this section, we develop a BCD-based algorithm for problem~\eqref{p1}. With the MA positions fixed, we optimize the beamforming vectors and AN covariance matrix using the S-procedure, SDR, and bisection search. We then update the auxiliary variables in closed form and optimize the MA positions via SCA.

\subsection{Optimizing Beamforming and AN Covariance Matrix}
For fixed MA positions, constraint \eqref{Const_1} can be reformulated as
\begin{equation}
\begin{aligned}
\frac{\bm h_k^H(\bm t)\bm W_k\bm h_k(\bm t)}
{\bm h_k^H(\bm t)(\bm R-\bm W_k)\bm h_k(\bm t)+\sigma_k^2}
\ge\nu_k\gamma
\end{aligned}
\label{eq:affine_sinr}
\end{equation}
By applying Lemma~1 in~\cite{ZhangRobustCovertISAC}, constraint~\eqref{Const_2} can be equivalently reformulated as the following LMI,
\begin{equation}
\bm\Theta_q=
\begin{bmatrix}
\omega_q\bm I_M-\bm\Xi&-\bm\Xi\bar{\bm h}_q(\bm t)\\
-\bar{\bm h}_q^H(\bm t)\bm\Xi&
\eta\sigma_q^2-\bar{\bm h}_q^H(\bm t)\bm\Xi\bar{\bm h}_q(\bm t)-\omega_q e_q^2
\end{bmatrix}\succeq\bm0,
\label{eq:Sproc_lmi}
\end{equation}
where $\bm\omega=[\omega_1,\ldots,\omega_Q]^T$ is the vector of nonnegative auxiliary variables,  and $\bm\Xi=(1+\eta)\sum_{k\in\mathcal K_c}\bm W_k-\eta\bm R$. For a given $\gamma>0$, the beamforming and AN covariance matrix subproblem is formulated as
\begin{subequations}\label{p2}
\begin{align}
\mathmakebox[2em][r]{\text{find}}\quad&\{\bm W_k\}_{k\in\mathcal K},\bm R,\bm\omega\\
\mathmakebox[2em][r]{\text{s.t.}}\quad&\frac{\bm h_k^H(\bm t)\bm W_k\bm h_k(\bm t)}
{\bm h_k^H(\bm t)(\bm R-\bm W_k)\bm h_k(\bm t)+\sigma_k^2}
\ge\nu_k\gamma,\ \forall k\in\mathcal K,\label{eq:p2_sinr}\\
&\bm\Theta_q(\bm t,\{\bm W_k\},\bm R,\omega_q)\succeq\bm0,\ \omega_q\ge0,\ \forall q\in\mathcal Q,\label{eq:p2_covert}\\
&\Tr(\bm R)\le P,\label{eq:p2_power}\\
&\bm R-\sum_{k\in\mathcal K}\bm W_k\succeq\bm0,\quad\bm W_k\succeq\bm0,\quad\forall k\in\mathcal K,\label{eq:p2_psd}\\
&\operatorname{rank}(\bm W_k)=1,\quad\forall k\in\mathcal K.\label{eq:p2_rank}
\end{align}
\end{subequations}
By dropping the rank-one constraints~\eqref{eq:p2_rank}, we obtain a relaxed version of~\eqref{p2}. When $\gamma$ and the covariances are jointly optimized,~\eqref{eq:affine_sinr} contains products of optimization variables. For fixed $\gamma$, this constraint is affine, and the relaxed problem is a semidefinite programming (SDP) feasibility problem. A solution feasible at $\gamma$ also remains feasible at any smaller nonnegative threshold. Thus, bisection can find the largest feasible threshold to a prescribed tolerance.

Let $\{\overline{\bm W}_k\}_{k\in\mathcal K}$ and $\overline{\bm R}$ denote a feasible solution to the relaxed version of~\eqref{p2} for a given $\gamma>0$.
A rank-one feasible solution preserving the SINR threshold can be recovered in closed form as
\begin{align}
&\bm R^{\star}= \overline{\bm R},
\bm w_k^\star=
(\bm h_k^H(\bm t)\overline{\bm W}_k\bm h_k(\bm t))^{-\frac{1}{2}}\overline{\bm W}_k\bm h_k(\bm t),\\
&\bm W_k^\star=\bm w_k^\star\bm w_k^{\star H},\bm R_{\rm A}^\star={\bm R^\star}-\sum_{k\in\mathcal K}\bm W_k^\star.
\label{SDR_optimal}
\end{align}

\subsection{Optimizing Auxiliary Variables}\label{subsec:auxiliary_update}
With other variables being fixed, we introduce an auxiliary variable $\alpha_k\in\mathbb C$ for each user. 
By applying the quadratic transform, the SINR in~\eqref{eq:sinr_user} can be recast as
\begin{align}
\tilde{\gamma}^{1}_{k}(\bm t,\!\alpha_k)
\!=\!2\Re\{\alpha_k^*\bm h_k^H\!(\bm t)\bm w_k\}
\!-\!|\alpha_k|^2\!\big(\bm h_k^H\!(\bm t)\bm\Xi_k
\bm h_k(\bm t)
\!+\!\sigma_k^2\big),
\label{eq:qt_position_sinr}
\end{align}
where $\bm\Xi_k= \bm R - \bm W_k$. 
Notably, for fixed antenna positions, the closed-form solution for $\alpha_k$ is given by
\begin{equation}
\alpha_k=
\frac{\bm h_k^H(\bm t)\bm w_k}
{\bm h_k^H(\bm t)\bm\Xi_k
\bm h_k(\bm t)+\sigma_k^2}, \forall k \in \mathcal{K}.
\label{eq:alpha_update_position}
\end{equation}

\subsection{Optimizing Transmit Antenna Positions}\label{subsec:MA_position_update_wc}
With the transmit beamforming vectors, AN covariance matrix, and auxiliary variables $\{\alpha_k\}_{k\in\mathcal K}$ fixed, the MA position subproblem can be formulated as
\begin{subequations}\label{eq:position_problem_fixed}
\begin{align}
\mathmakebox[2em][r]{\max_{\mathclap{\bm t,\gamma}}}\quad&\gamma\\
\mathmakebox[2em][r]{\text{s.t.}}\quad&\gamma^{1}_{k}(\bm t)\ge\nu_k\gamma,\forall k\in\mathcal K,\label{eq:position_sinr_constraint}\\
&\bm h_q^H\!(\bm t)\bm\Xi\bm h_q(\bm t)\!\le\!\eta\sigma_q^2,\norm{\bm\delta_q}_2\le e_q, \forall q\in\mathcal Q,\label{eq:position_covert_constraint}\\
&t_1\ge0,t_M\le D,\\
&t_m-t_{m-1}\ge d,2\le m\le M.
\end{align}
\end{subequations}
Note that problem~\eqref{p1} is intractable since the optimization variables in the SINR and covertness constraints are strongly coupled.
To tackle the intractable subproblem, we employ SCA to construct a sequence of convex subproblems.

To approximate the two terms involving $\bm t$ in~\eqref{eq:qt_position_sinr} separately, we define $a_k(\bm t)$ and $b_k(\bm t)$ as
\begin{equation}\label{eq:position_ab_definitions}
\begin{aligned}
a_k(\bm t)\!=\!2\Re\{\alpha_k^*\bm h_k^H\!(\bm t)\bm w_k\},
b_k(\bm t)\!=\!|\alpha_k|^2\bm h_k^H\!(\bm t)\bm\Xi_k\bm h_k(\bm t),
\end{aligned}
\end{equation}
Note that $a_k(\bm t)$ and $b_k(\bm t)$ contain cosine terms and are generally neither convex nor concave with respect to (w.r.t.) $\bm t$.
To address this challenge, we construct a concave lower bound for $a_k(\bm t)$ and a convex upper bound for $b_k(\bm t)$ using second-order Taylor bounds, i.e.,
\begin{equation}\label{eq:position_taylor_bounds}
\begin{aligned}
&a_k(\bm t)\!\ge\! a_k(\bm t_0)\!+\!\nabla a_k(\bm t_0)^T(\bm t\!-\!\bm t_0)\!-\!\frac{\varpi_{a,k}}{2}\|\bm t\!-\!\bm t_0\|_2^2
\!\triangleq\!\underline a_k(\bm t),\\
&b_k(\bm t)\!\le\! b_k(\bm t_0)\!+\!\nabla b_k(\bm t_0)^T(\bm t\!-\!\bm t_0)\!+\!\frac{\varpi_{b,k}}{2}\|\bm t\!-\!\bm t_0\|_2^2
\!\triangleq\!\bar b_k(\bm t),
\end{aligned}
\end{equation}
where $\bm t_0$ denotes the expansion point at the current iteration. The vectors $\nabla a_k(\bm t)$ and $\nabla b_k(\bm t)$ denote the gradients of $a_k(\bm t)$ and $b_k(\bm t)$ w.r.t.\ $\bm t$, respectively, and $\nabla^2a_k(\bm t)$ and $\nabla^2b_k(\bm t)$ are the corresponding Hessian matrices. The nonnegative constants $\varpi_{a,k}$ and $\varpi_{b,k}$ satisfy $\varpi_{a,k}\ge\|\nabla^2a_k(\bm t)\|_2$ and $\varpi_{b,k}\ge\|\nabla^2b_k(\bm t)\|_2$.
By applying the Gershgorin circle theorem~\cite{HornJohnson} to the Hessian matrices, we can obtain
$\varpi_{a,k}=\frac{8\pi^2|\alpha_k|}{\lambda^2}
\sum_{\ell=1}^{L}|\beta_k^\ell|\cos^2\theta_k^\ell\max_m|[\bm w_k]_m|$
and $\varpi_{b,k}=\frac{16\pi^2M|\alpha_k|^2}{\lambda^2}
\sum_{\ell=1}^{L}\sum_{p=1}^{L}|\beta_k^\ell||\beta_k^p|
\max_{m,n}|[\bm\Xi_k]_{m,n}|$.
Substituting \eqref{eq:position_taylor_bounds} into~\eqref{eq:qt_position_sinr}, we obtain
$ \underline{\gamma}^{1}_{k}(\bm t)
=\underline a_k(\bm t)-\bar b_k(\bm t)-|\alpha_k|^2\sigma_k^2.$
The resulting conservative convex constraint is
\begin{equation}\label{eq:sinr_sca_constraint_position}
\underline{\gamma}^{1}_{k}(\bm t)\ge\nu_k\gamma,\quad\forall k\in\mathcal K.
\end{equation}

Next, we construct a convex approximation of the covertness constraint~\eqref{eq:position_covert_constraint}. With~\eqref{eq:warden_CSI_decomposition} substituted into its left-hand side (LHS), the covertness constraint~\eqref{eq:position_covert_constraint} can be recast as
$\bm\delta_q^H\bm\Xi\bm\delta_q+\mu_q(\bm t,\bm\delta_q)+\varkappa_q(\bm t)\le\eta\sigma_q^2$,
where $\mu_q(\bm t,\bm\delta_q)\triangleq\bar{\bm h}_q^H(\bm t)\bm\Xi\bm\delta_q+\bm\delta_q^H\bm\Xi\bar{\bm h}_q(\bm t)$ and $\varkappa_q(\bm t)\triangleq\bar{\bm h}_q^H(\bm t)\bm\Xi\bar{\bm h}_q(\bm t)$.
we recast $\mu_q(\bm t,\bm\delta_q)$ as
$\mu_q(\bm t,\bm\delta_q)
=\sum_{m=1}^{M}
\bigl(\bm p_{q,m}(t_m)\bm\delta_q
+\bm\delta_q^H\bm p_{q,m}^H(t_m)\bigr)$,
where $\bm p_{q,m}(t_m)\triangleq\beta_q e^{\jmath\frac{2\pi}{\lambda}t_m\cos\theta_q}[\bm\Xi]_{m,:}$ and $[\bm\Xi]_{m,:}$ denotes the $m$-th row of $\bm\Xi$.
The $m$-th element of $\nabla\mu_q(\bm t,\bm\delta_q)$ can be expressed as
$\frac{\partial\mu_q(\bm t,\bm\delta_q)}{\partial t_m}
=\bm r_{q,m}(t_m)\bm\delta_q
+\bm\delta_q^H\bm r_{q,m}^H(t_m)$,
where $\bm r_{q,m}(t_m)\triangleq\jmath\frac{2\pi}{\lambda}\rho_q\bm p_{q,m}(t_m)$ and $\rho_q=\cos\theta_q$.
Then, we have
$\frac{\partial^2\mu_q(\bm t,\bm\delta_q)}{\partial t_m^2}
=-\frac{4\pi^2\rho_q^2}{\lambda^2}
\bigl(\bm p_{q,m}(t_m)\bm\delta_q
+\bm\delta_q^H\bm p_{q,m}^H(t_m)\bigr)$ and
$\frac{\partial^2\mu_q(\bm t,\bm\delta_q)}{\partial t_m\partial t_n}=0$ for $m\ne n$.
By applying the triangle and Cauchy--Schwarz inequalities, we have$
\left|\frac{\partial^2\mu_q(\bm t,\bm\delta_q)}{\partial t_m^2}\right|
\!\le\!\frac{8\pi^2\rho_q^2|\beta_q|e_q}{\lambda^2}
\max_n\|[\bm\Xi]_{n,:}\|_2\triangleq\varpi_{\mu,q},\!
$
where $\varpi_{\mu,q}\bm I_M\succeq\nabla^2\mu_q(\bm t,\bm\delta_q)$.
Therefore, we have $\mu_q(\bm t,\bm\delta_q)\le\bar\mu_q(\bm t,\bm\delta_q)$,
where $\bar\mu_q(\bm t,\bm\delta_q)\triangleq\mu_q(\bm t_0,\bm\delta_q)
+\nabla\mu_q(\bm t_0,\bm\delta_q)^T(\bm t-\bm t_0)
+\frac{\varpi_{\mu,q}}{2}\|\bm t-\bm t_0\|_2^2$.
Similarly, we have $\varkappa_q(\bm t)\le\bar\varkappa_q(\bm t)$,
where $\bar\varkappa_q(\bm t)\triangleq\varkappa_q(\bm t_0)
+\nabla\varkappa_q(\bm t_0)^T(\bm t-\bm t_0)
+\frac{\varpi_{\varkappa,q}}{2}\|\bm t-\bm t_0\|_2^2$.
The corresponding Hessian bound is given by
$
\varpi_{\varkappa,q}
=\frac{16\pi^2(M-1)|\beta_q|^2}{\lambda^2}
\max_{m,n}|[\bm\Xi]_{m,n}|,
$
where $\varpi_{\varkappa,q}\bm I_M\succeq\nabla^2\varkappa_q(\bm t)$.
Then, substituting $\bar\mu_q(\bm t,\bm\delta_q)$ and $\bar\varkappa_q(\bm t)$ into the reformulated covertness constraint, we obtain
\begin{equation}\label{eq:covert_sca_expanded_position}
\begin{aligned}
\bm\delta_q^H\bm\Xi\bm\delta_q
+\bar\mu_q(\bm t,\bm\delta_q)
+\bar\varkappa_q(\bm t)-\eta\sigma_q^2\le0,\|\bm\delta_q\|_2\le e_q,\forall q \in \mathcal{Q}.
\end{aligned}
\end{equation}
We rewrite $\nabla\mu_q(\bm t_0,\bm\delta_q)^T(\bm t-\bm t_0)$ as
$\nabla\mu_q(\bm t_0,\bm\delta_q)^T(\bm t-\bm t_0)
=\bm z_q(\bm t)^H\bm\delta_q
+\bm\delta_q^H\bm z_q(\bm t),$
where
$\bm z_q(\bm t)
\triangleq\bigl[\bm r_{q,1}^H(\bm t_0^1),\ldots,\allowbreak
\bm r_{q,M}^H(\bm t_0^M)\bigr](\bm t-\bm t_0).$
Hence,~\eqref{eq:covert_sca_expanded_position} can be written in the following tractable form, i.e.
\begin{equation}\label{eq:position_covert_majorization}
\begin{aligned}
&\bm\delta_q^H\bm\Xi\bm\delta_q
+\bm\delta_q^H\bm c_q(\bm t)
+\bm c_q(\bm t)^H\bm\delta_q+\upsilon_q(\bm t)-\eta\sigma_q^2\le0,
\end{aligned}
\end{equation}
where $\bm c_q(\bm t)\triangleq\bm\Xi\bar{\bm h}_q(\bm t_0)+\bm z_q(\bm t)$ and $\upsilon_q(\bm t)=\varkappa_q(\bm t_0)+\nabla\varkappa_q(\bm t_0)^T(\bm t-\bm t_0)+\frac{\varpi_{\varkappa,q}+\varpi_{\mu,q}}{2}\|\bm t-\bm t_0\|_2^2$.
By the S-procedure~\cite{ZhangRobustCovertISAC},~\eqref{eq:position_covert_majorization} is equivalent to the existence of $\omega_q\ge0$ such that
\begin{equation}\label{eq:position_covert_sca_matrix}
\begin{bmatrix}
\omega_q\bm I_M-\bm\Xi&-\bm c_q(\bm t)\\
-\bm c_q(\bm t)^H&\eta\sigma_q^2-\upsilon_q(\bm t)-\omega_qe_q^2
\end{bmatrix}\succeq\bm0.
\end{equation}
However, the matrix inequality in~\eqref{eq:position_covert_sca_matrix} is still nonlinear w.r.t.\ $\bm t$ due to the quadratic term in $\upsilon_q(\bm t)$.
To obtain an LMI formulation, we introduce the following lemma.
\begin{lemma}\phantomsection\label{lem:lmi_brc_relaxation}
\it For any complex Hermitian matrix $\boldsymbol A\in\mathbb{C}^{N\times N}$, $\boldsymbol b\in\mathbb{C}^{N\times 1}$, $c\in\mathbb R$, and $d\in\mathbb R$, if
\begin{equation}\label{eq:brc_psd_condition_position}
\begin{bmatrix}
\boldsymbol A & \boldsymbol b\\
\boldsymbol b^H & c
\end{bmatrix}\succeq\boldsymbol 0
\quad\text{and}\quad d\ge c,
\end{equation}
then
\begin{equation}\label{eq:brc_psd_implication_position}
\begin{bmatrix}
\boldsymbol A & \boldsymbol b\\
\boldsymbol b^H & d
\end{bmatrix}\succeq\boldsymbol 0
\end{equation}
\end{lemma}

By applying Lemma~\ref{lem:lmi_brc_relaxation}, the matrix inequality~\eqref{eq:position_covert_sca_matrix} is equivalently reformulated as the following LMI together with the auxiliary constraint~\eqref{eq:iota_position},
\begin{equation}\label{eq:position_covert_sca_lmi_relaxed}
\bm\Theta_{q,t}(\bm t,\omega_q,\iota_q)\!\triangleq\!
\begin{bmatrix}
\omega_q\bm I_M-\bm\Xi&-\bm c_q(\bm t)\\
-\bm c_q(\bm t)^H&\eta\sigma_q^2-\iota_q-\omega_qe_q^2
\end{bmatrix}\succeq\bm0,
\end{equation}
where $\iota_q\in\mathbb R$ is an auxiliary variable satisfying
\begin{equation}\label{eq:iota_position}
\iota_q\ge\upsilon_q(\bm t).
\end{equation}

Based on the above derivations, a convex approximation of problem~\eqref{eq:position_problem_fixed} is formulated as
\begin{equation}\label{p3_sca}
\begin{aligned}
\mathmakebox[2em][r]{\max_{\mathclap{\bm t,\gamma,\bm\omega,\bm\iota}}}\quad&\gamma\\
\mathmakebox[2em][r]{\text{s.t.}}\quad&\eqref{eq:sinr_sca_constraint_position},~\eqref{eq:position_covert_sca_lmi_relaxed},~\eqref{eq:iota_position},~\eqref{Const_4},~\eqref{Const_5},
\end{aligned}
\end{equation}
where $\bm\omega=[\omega_1,\ldots,\omega_Q]^T\in\mathbb R^Q$, and $\bm\iota=[\iota_1,\ldots,\iota_Q]^T\in\mathbb R^Q$.  Problem~\eqref{p3_sca} is convex and can be solved using CVX toolbox.

\subsection{Algorithm Framework and Complexity Analysis}\label{subsec:algorithm_complexity}
The overall BCD procedure is summarized in Algorithm~\ref{alg:robust_bcd}.
We next analyze the complexity of the convex subproblems using the LMI and SOC counting method in~\cite{WangOutageConic2014}. The beamforming and AN covariance matrix subproblem has complexity $\mathcal C_{{\rm B},{\rm WC}}=\mathcal O\bigl({(K+Q)^{1.5}}\allowbreak M^{2.5}\allowbreak{(KM^2+Q)^2}\bigr)$ for the worst-case optimization scheme. The MA position subproblem has complexity $\mathcal C_{{\rm X},{\rm WC}}=\mathcal O\bigl(M^2(M+2Q)\allowbreak(QM+K)^{0.5}\allowbreak(QM+2Q^2+K)\bigr)$ for the worst-case optimization scheme.

\begin{algorithm}[!t]
\small
\SetAlgoNoLine
\SetNlSty{textnormal}{}{:\,}
\SetAlgoNlRelativeSize{0}
\SetAlgoCaptionSeparator{}
\SetAlCapHSkip{0pt}
\SetAlgoHangIndent{0pt}
\SetArgSty{textnormal}
\SetKwInput{KwInit}{Initialize}
\caption{BCD Algorithm for Solving Problem \eqref{p1}}
\label{alg:robust_bcd}
\KwInit{$\bm t^{(r)}$, $\{\boldsymbol w_k^{(r)}\}$, $\boldsymbol R_{\rm A}^{(r)}$, and set $r=0$.}

\SetEndCharOfAlgoLine{.}
\Repeat{exit conditions are met}{
    \SetEndCharOfAlgoLine{;}
    Update $\gamma_{\rm B}^{(r+1)}$ by bisection search\;
    Update $\overline{\boldsymbol W}_k^{(r+1)}$ and $\overline{\boldsymbol R}^{(r+1)}$ via \eqref{p2} at $\gamma_{\rm B}^{(r+1)}$\; 
    Construct $\boldsymbol w_k^{(r+1)}$ and $\boldsymbol R_{\rm A}^{(r+1)}$ via \eqref{SDR_optimal}\;
    Update $\alpha_k^{(r+1)}$ via \eqref{eq:alpha_update_position}\;
    Update $\bm t^{(r+1)}$ and $\gamma^{(r+1)}$ via \eqref{p3_sca}\;
    Update $r=r+1$\;
}
\KwRet{$\bm t^\star$, $\boldsymbol w_k^\star$, and $\boldsymbol R_{\rm A}^\star$}\;
\end{algorithm}
\begin{figure*}[t]
\captionsetup[subfloat]{font=footnotesize}
\subfloat[Convergence of the proposed algorithms.]{%
\makebox[0.33\textwidth][c]{%
\includegraphics[width=0.33\textwidth]{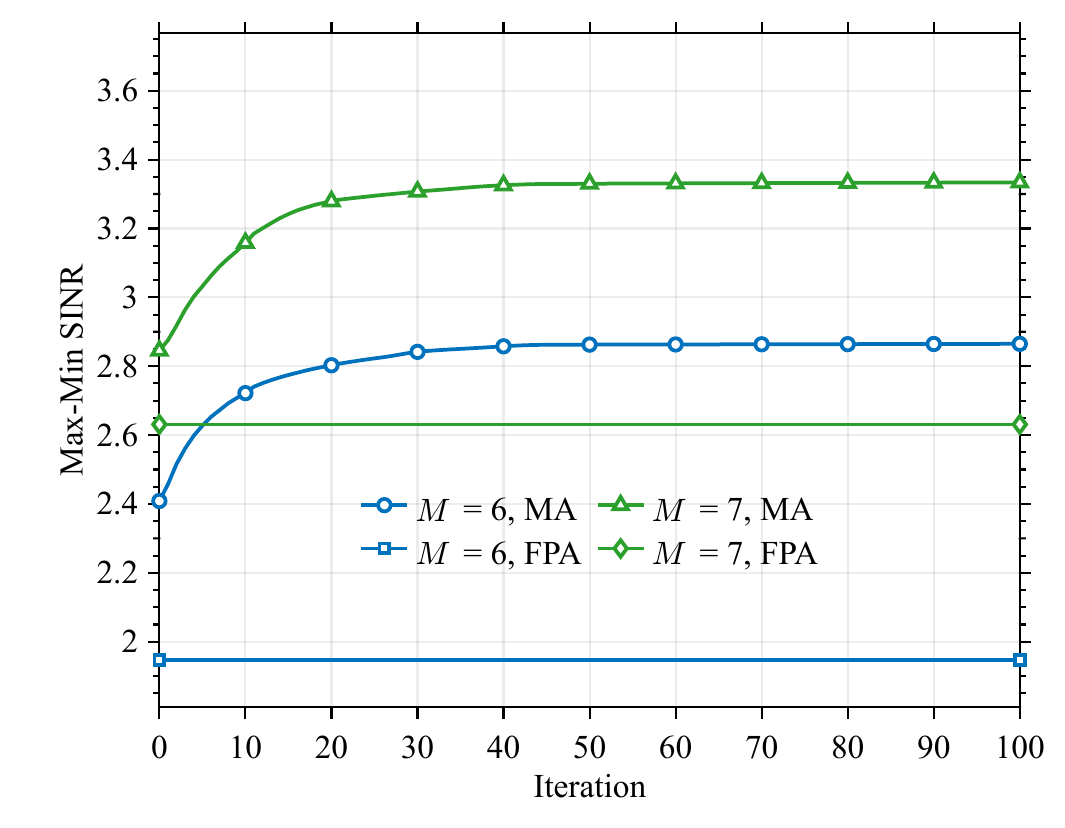}}\label{fig:iteration}}\hfill
\subfloat[Weighted max-min SINR versus total power.]{%
\makebox[0.33\textwidth][c]{%
\includegraphics[width=0.33\textwidth]{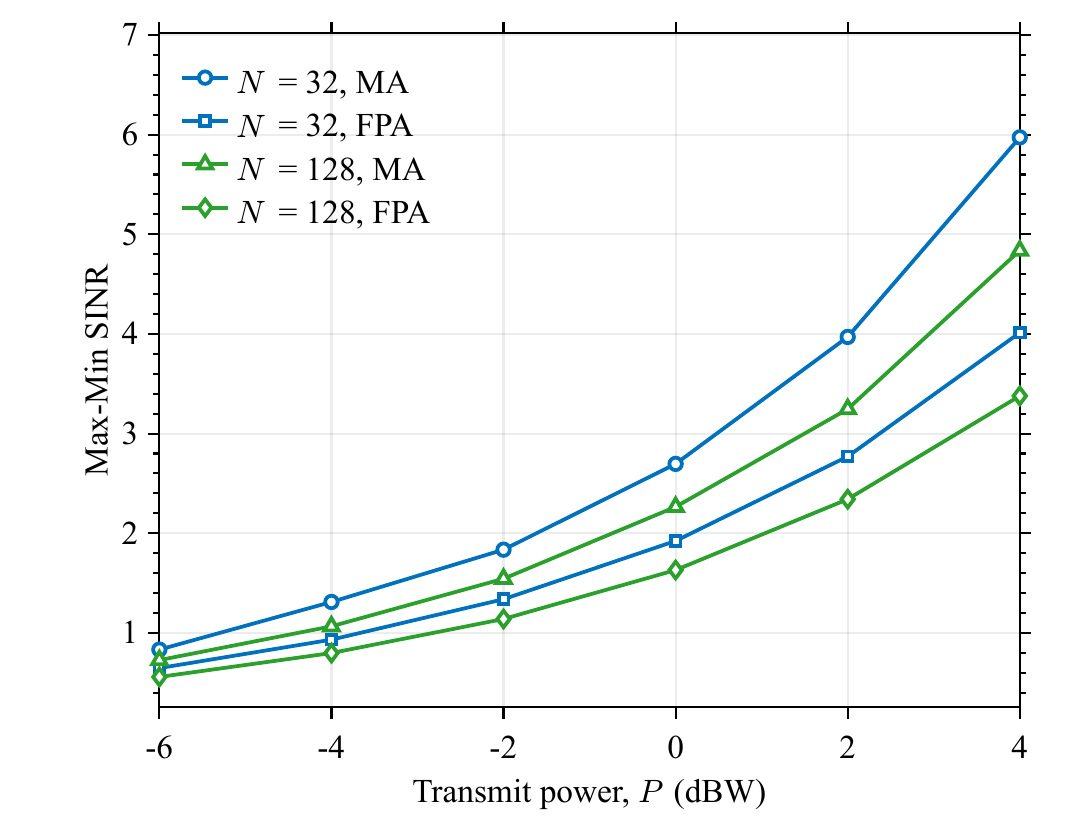}}\label{fig:power}}\hfill
\subfloat[Weighted max-min SINR versus $\kappa_{\mathrm{CSI}}$.]{%
\makebox[0.33\textwidth][c]{%
\includegraphics[width=0.33\textwidth]{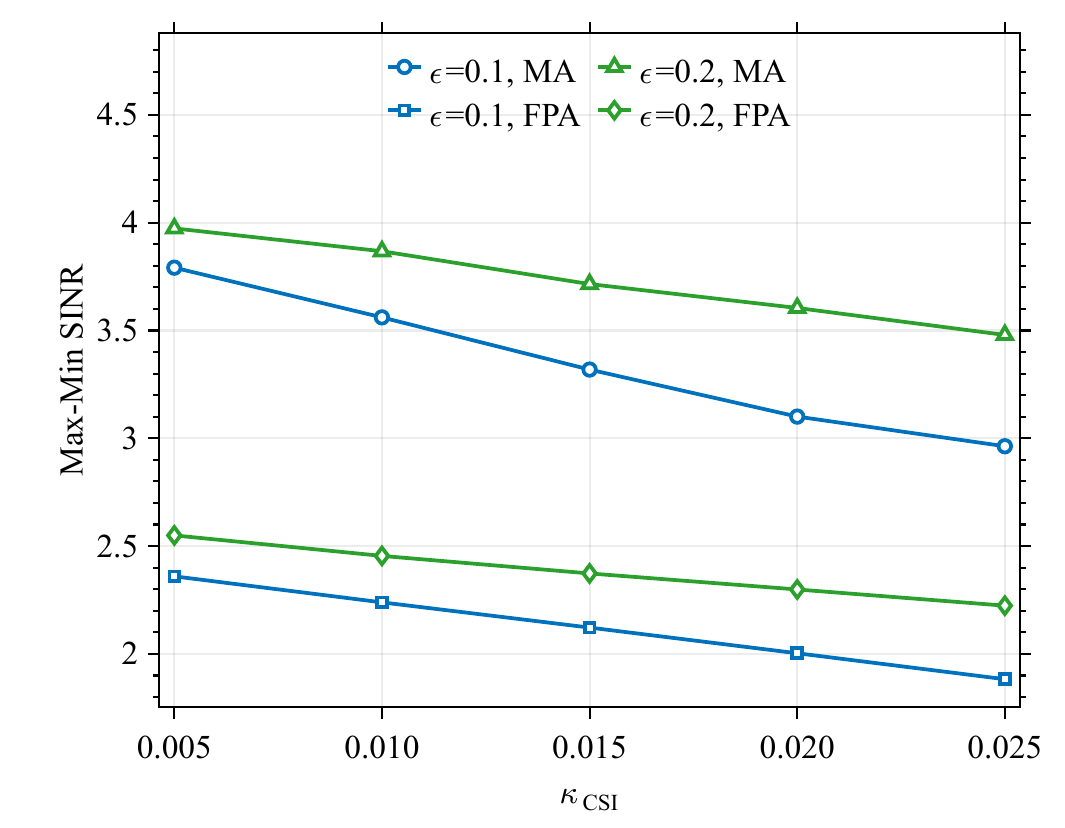}}\label{fig:csi_error}}
\caption{Performance comparison of different schemes.}
\label{fig:results}
\end{figure*}
\section{Simulation Results}
This section presents numerical results to evaluate the performance of the proposed MA-assisted covert communication scheme. Unless otherwise specified, $M=6$, $K_o=K_c=Q=2$, $P=0$~dBW, and the noise power at all users and Willies is $-80$~dBW. Alice is located at $(0,0)\rm m$. Overt users 1 and 2 are randomly placed in disks of radius $r_o=15$~m centered $60$~m from Alice at $70^\circ$ and $110^\circ$, respectively.
One covert user is located at the center of each disk.
The Willies are located $60$~m from Alice, and the azimuth angles from Alice to the two Willies are $80^\circ$ and $100^\circ$, respectively. 
The linear transmit region has length $D=12\lambda$, with $\lambda=0.1$~m and minimum MA spacing $d=\lambda/2$. Each user channel contains one LoS path and $L-1$ non-line-of-sight (NLoS) paths, with $L=3$. The independent path gains satisfy $\beta_k^1\sim\mathcal{CN}\bigl(0,{\kappa c_k^2}/(\kappa+1)\bigr)$ for the LoS path and $\beta_k^\ell\sim\mathcal{CN}\bigl(0,{c_k^2}/{((\kappa+1)(L-1))}\bigr)$ for the NLoS paths, where $\ell=2,\ldots,L$ and $\kappa=2$. The large-scale path loss is $c_k^2=C_0d_k^{-\alpha_{\mathrm{PL}}}$, where $d_k$ is the link distance in meters, $C_0=10^{-3}$, and $\alpha_{\mathrm{PL}}=2.6$. Similarly, the Willie channel gain is $\beta_q=c_qe^{\jmath\phi_q}$, where $\phi_q\sim\mathcal{U}[0,2\pi)$, $c_q^2=C_0d_q^{-\alpha_{\mathrm{PL}}}$, and $d_q$ is the link distance in meters. The LoS azimuth angles are determined by the user and Willie locations, while the NLoS azimuth angles are independently and uniformly distributed over $[0,\pi]$. The transmission blocklength is $N=32$, the covertness level is $\epsilon=0.1$, and the covert user priority parameter is $\nu=0.5$. The CSI uncertainty coefficient is set to $\kappa_{\mathrm{CSI}}=0.01$. The bounded-error radius is chosen as $e_q^2=\kappa_{\mathrm{CSI}}\norm{\bar{\bm h}_q(\bm t)}_2^2$. 

Fig.~\ref{fig:results}\subref{fig:iteration} depicts the convergence behavior of the proposed MA algorithms for $M=6$ and $7$. It can be observed that the weighted max-min SINR achieved by the proposed MA algorithms increases monotonically and converges rapidly after 30 iterations.
In addition, from Fig.~\ref{fig:results}\subref{fig:iteration}, we can observe that the weighted max-min SINR improves with increasing $M$.
The reason lies in the additional DoFs provided by more antenna elements.

Fig.~\ref{fig:results}\subref{fig:power} illustrates the weighted max-min SINR versus the maximum transmit power $P$ for different transmission blocklengths.
It can be observed that the weighted max-min SINRs of all schemes increase as $P$ increases.
This is because a larger transmit-power budget provides more power for the transmit signal while satisfying the covertness constraints. Moreover, the proposed MA schemes outperform the conventional FPA schemes by properly optimizing the antenna positions to enhance the desired channels and mitigate interference. It can also be observed that the weighted max-min SINR achieved with $N=128$ is lower than that with $N=32$.
This is because a larger $N$ provides each Willie with more observations, improving the accuracy of received power estimation and the reliability of covert transmission detection.
For a fixed received power ratio, the KL divergence in \eqref{eq:KL_div} increases linearly with $N$.
For a fixed $\epsilon$, the threshold $\eta$ therefore decreases, imposing more stringent constraints on covert signal leakage.
These constraints restrict the feasible region and limit the achievable weighted max-min SINR.

In Fig.~\ref{fig:results}\subref{fig:csi_error}, we present the weighted max-min SINR versus the relative CSI uncertainty $\kappa_{\mathrm{CSI}}$. It can be observed that the weighted max-min SINR of all schemes decreases with increasing $\kappa_{\mathrm{CSI}}$. This is because increased uncertainty in the Willie channels makes it more difficult to satisfy the covertness constraints. For a fixed $\epsilon$, the proposed MA schemes outperform the conventional FPA schemes by properly optimizing the antenna positions to enhance the desired user channels and suppress information leakage.
\section{Conclusion}
In this paper, we investigated MA-assisted multiuser covert communications with multiple Willies under norm-bounded CSI errors.
We aimed to maximize the weighted minimum SINR by jointly optimizing the transmit beamforming vectors, AN covariance matrix, and MA positions subject to worst-case covertness constraints.
To solve this problem, we developed a BCD-based algorithm incorporating the S-procedure, SDR, bisection search, and SCA methods.
Simulation results demonstrated that the proposed MA scheme outperforms the FPA schemes in terms of weighted max-min SINR.
\appendices

\bibliographystyle{IEEEtran}
\bibliography{reference}
\end{document}